\documentclass[mathleft
]{an}
\usepackage{graphicx,epsfig,subfig}
\usepackage{times}

\usepackage{natbib}
\bibpunct{(}{)}{;}{a}{}{,}
\begin{document}


\title{Study of resonances for the restricted 3-body problem}

\author{B. Quarles\inst{1},
Z.E. Musielak\inst{1,2}
\and  M. Cuntz\inst{1} \fnmsep\thanks{Corresponding author:
  \email{cuntz@uta.edu}\newline}
}
\titlerunning{Instructions for authors}
\authorrunning{T.H.E. Editor \& G.H. Ostwriter}
\institute{
Department of Physics, Science Hall, University of Texas at Arlington,
Arlington, TX 76019-0059, USA
\and 
Kiepenheuer-Institut f\"ur Sonnenphysik, Sch\"oneckstr. 6, D-79104 Freiburg, Germany
}

\received{xx Nov 2011}
\accepted{xx xxx xxxx}
\publonline{later}

\keywords{stars: binaries: general -- celestial mechanics -- chaos -- stars: planetary systems}

\abstract{%
Our aim is to identify and classify mean-motion resonances (MMRs) for the coplanar
circular restricted three-body problem (CR3BP) for mass ratios between
0.10 and 0.50.  Our methods include the maximum Lyapunov exponent,
which is used as an indicator for the location of the resonances,
the Fast Fourier Transform (FFT) used for determining what
kind of resonances are present, and the inspection of the orbital elements
to classify the periodicity. We show that the 2:1 resonance occurs
the most frequently.  Among other resonances, the 3:1 resonance is
the second most common, and furthermore both 3:2 and 5:3 resonances
occur more often than the 4:1 resonance.  Moreover, the resonances 
in the coplanar CR3BP are classified based on the behaviour of the orbits.
We show that orbital stability is ensured for high values of resonance
(i.e., high ratios) where only a single resonance is present.  The
resonances attained are consistent with the previously established
resonances for the Solar System, i.e., specifically, in regards to
the asteroid belt.  Previous work employed digital filtering and
Lyapunov characteristic exponents to determine stochasticity of the
eccentricity, which is found to be consistent with our usage of Lyapunov
exponents as an alternate approach based on varying the mass ratio instead
of the eccentricity.  Our results are expected to be of principal interest
to future studies, including augmentations to observed or proposed resonances,
of extra-solar planets in binary stellar systems.
}

\maketitle

\section{Introduction} \label{intro}

The circular restricted three-body problem (CR3BP) has been considered in
orbital mechanics for more than two centuries, and its numerous applications to 
describe the motion of planets, asteroids and artificial satellites in the 
Solar System are well-known
\citep[][~and references therein]{sze67,dvo84,dvo86,rab88,smi93,mur00}.  
More recently, the study of the CR3BP has been motivated as a testbed for
investigations of the orbital stability of extra-solar planetary systems
around single stars as well as planets in stellar binary systems.

Extensive studies of orbital stability in the CR3BP have been
performed by many authors, who used tools ranging from the
Jacobi constant \citep*[e.g.,][]{cun07,ebe08},
a hodograph-based method \citep{ebe10} and Lyapunov exponents
\citep*[e.g.,][]{gon81,jef83,lec92,mil92,smi93,lis99,mur01,qua11}.
There have also been inquiries into different numerical methods
specifically designed to study the orbital stability problem
\citep[e.g.,][]{yos90,hol99,dav03,mus05}.
Some recent studies involve spacecraft trajectories to
the Lagrangian point L$_3$ \citep{tar10} and the stability analysis
of artificial equilibrium points in the CR3BP \citep{bom11}.    

The origin and nature of mean-motion resonances (MMRs) in the CR3BP,
particularly for cases of low mass ratios,
have also been studied, and the role of MMRs in orbital stability has been 
investigated in detail.  \cite{had93} identified three 
(2:1, 3:1 and 4:1) main MMRs in the Sun-Jupiter and asteroid system.
Extensive studies of MMRs, with extensions to the elliptic restricted
three-body problem (ER3BP), have been performed by \cite{fer94,nes02,pil02,hag03,vel04,mar07,sze08}.  
It has been shown that the main difference in orbital stability 
between stable and unstable systems with MMRs is that the latter involves 
more than one resonance, and that the existence of two or more resonances 
leads to their overlap making some orbits unstable \citep[e.g.,][]{mud06,mar07}.

This type of work is also motivated by earlier results on
various extra-solar star--planet systems.  Previously, \cite{mar01} and
\cite{lee06} discovered two planets in 2:1 orbital resonance about
the star GJ~876 and HD~82943, respectively.  Other interesting cases
include the 3:1 MMR in the five planet system of 55~Cnc, occurring
between the inner planets 55~Cnc~c and 55~Cnc~b \citep*{mar02,ji03,nov03}
and the 3:2 MMR in the system of HD~45364 \citep{cor09}.
Furthermore, the two outer planets of 47~UMa are found to be
in a 5:2 resonance \citep{fis02}.  This system
is of particular interest as it is hosting two gas planets
with a similar mass ratio than that between Jupiter and Saturn,
which are orbiting a star of similar spectra type than the Sun.
However, the gas planets in this system are relatively close to
the zone of habitability; therefore, they adversely affect the
astrobiological significance of the system \citep*{nob02,goz02,cun03}.

To identify MMRs in the CR3BP, particularly for small mass ratios,
or in the elliptical restricted three-body problem (ER3BP)
several different methods have been developed.  \cite{had93} used 
the method of averaging as well as the method of computing periodic orbits 
numerically.  He showed how to utilize these methods to determine MMRs;
furthermore, he discussed the advantages and disadvantages of each method as well as the effects of 
their inherent limitations on the attained results.  Another method of finding MMRs 
is based on the concept of Lyapunov exponent as introduced by 
\cite{nes98} with its further development by \cite{mor99} and \cite{nes02}.  
So far, the developed methods have mainly been applied to identify MMRs that
affect the structure and evolution of the asteroid belt.  This means that the 
identified MMRs were found for the specific mass ratios of the Solar System primaries 
(Sun and Jupiter, and Jupiter and Saturn).  However, there is significant
ongoing work as well as the need for future work pertaining to extra-solar
planetary systems, which is self-evident from the recent discoveries by 
the Kepler space telescope of the circumbinary planets Kepler-16, Kepler-34
and Kepler-35 \citep{doy11,wel12}.

The main purpose of this paper is to identify and classify MMRs in the 
CR3BP for a broad range of mass ratios of the stellar components.  Our method
of finding resonances is based on the maximum Lyapunov exponent, which is 
used as an indicator for the location of resonances once the masses of the 
two massive bodies are specified.  Then, a Fast Fourier Transform (FFT) 
is used to determine what kind of resonances exist.  Thereafter, periodicity
is determined through the inspection 
of the orbit diagrams and osculating semimajor axis of representative cases.  Future 
applications of enhanced versions of our results will involve (1) extra-solar planetary systems
around single stars (2) exomoons in star--planet systems, and (3) planets
in stellar binary systems.  Currently, the ideal application of the CR3BP
to planets in stellar binary systems remains undetected.  
Therefore, specific applications are beyond
the scope of this paper that is fully devoted to the problem of finding
and classifying MMRs in the CR3BP.  Nonetheless, our present study is an
important step toward envisioned future investigations motivated by
existing and anticipated discoveries of planets through ongoing and future
search missions.

The paper is structured as follows:  Our method is described in Sect.~2
and in Sect.~3 we present our discussion and results.  Our summary and
conclusions are given in Sect.~4.


\section{Theorectical Approach} \label{sec:2}

We consider the CR3BP with objects of masses $m_1$, $m_2$ and $m_3$; note that
mass $m_3$ is assumed to be negligible compared to $m_1$ and $m_2$.  From 
a mathematical point of view, the problem is described by the following 
set of equations \citep{sze67}
\begin{equation} 
\ddot{x} - 2\dot{y} = {\partial{\Omega} \over \partial{x}}, \ \ 
\ddot{y} + 2\dot{x} = {\partial{\Omega} \over \partial{y}}, \ \
\ddot{z} = {\partial{\Omega} \over \partial{z}},
\label{eq1}
\end{equation}

\noindent
where the potential function $\Omega$ is given by

\begin{equation}
\Omega \ = \ {\alpha \over r_1} + {\mu \over r_2} + {1\over 2}
\left(\alpha r_1^2 + \mu r_2^2\right)
\end{equation}

\noindent
with $\mu = m_2 / (m_1 + m_2)$, $\alpha = 1 - \mu$, and
\begin{eqnarray}
r_1^2 \ & = & \ \left(x-\mu\right)^2 + y^2 + z^2     \\    \nonumber
r_2^2 \ & = & \ \left(x+\alpha\right)^2 + y^2 + z^2  \ .   \nonumber
\end{eqnarray}

\noindent
The Jacobian integral of Eq. (\ref{eq1}) is 
\begin{equation}
C \ =  \ 2\Omega - \left(\dot{x}^2 + \dot{y}^2 + \dot{z}^2\right)
\end{equation}
\noindent
where $C$ is the so-called Jacobi constant; see also previous papers
of this series for additional explanation on the adopted method. 

We study the manifestations of resonances in numerical solutions in the 
physical plane ($\textit{x}$, $\textit{y}$) for the Copenhagen case $\left(\mu = 0.5\right)$ 
and the special cases $\mu = 0.1$, $0.2$, $0.3$, and $0.4$.  We vary the 
parameter $\rho_0$, which is defined by the ratio of the initial distance 
of mass $m_3$ from the host mass $m_1$, given as $R_0$, and the separation distance,
$D$, between the two larger ($m_1$ and $m_2$) masses, i.e., $\rho_0=R_0/D$.
As an initial condition, the object of mass $m_3$ starts at the 3 o'clock
position with respect to the massive host star, $m_1$.  We also assume a
circular initial velocity for $m_3$.  The numerical simulations are performed
using a sixth order symplectic integration method with a constant time step
of 10$^{-3}$ binary periods per step \citep{yos90}.  Unstable systems are 
confirmed using a sixteenth order Gragg-Burlisch-Stoer integration method 
with the aforementioned constant time step as an initial guess \citep{gra96}.  
Additionally, we use fitting formulas \citep{hol99} to confirm the 
numerically determined stability limits.

\begin{figure*}[ht]

  \centering
  \epsfig{file=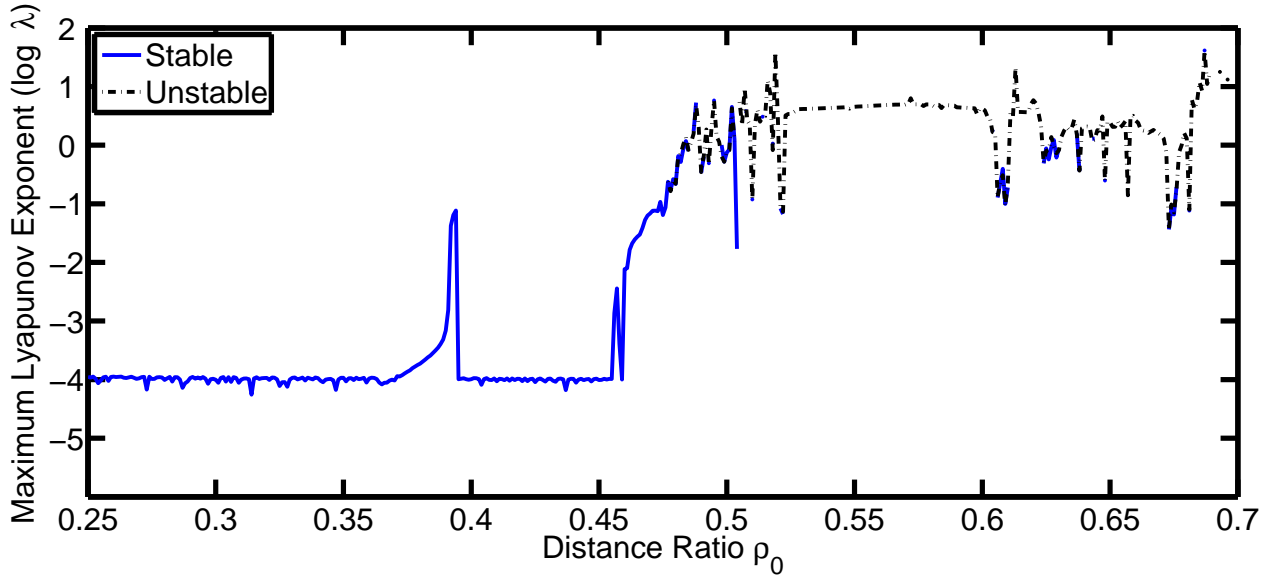,width=1.0\linewidth}

\caption{Maximum Lyapunov exponent is plotted versus the distance ratio 
$\rho_0$ for $\mu = 0.1$.  The stable and unstable cases are denoted.}
\label{fig:1}       
\end{figure*}

\begin{figure*}[ht]
\centering
  \epsfig{file=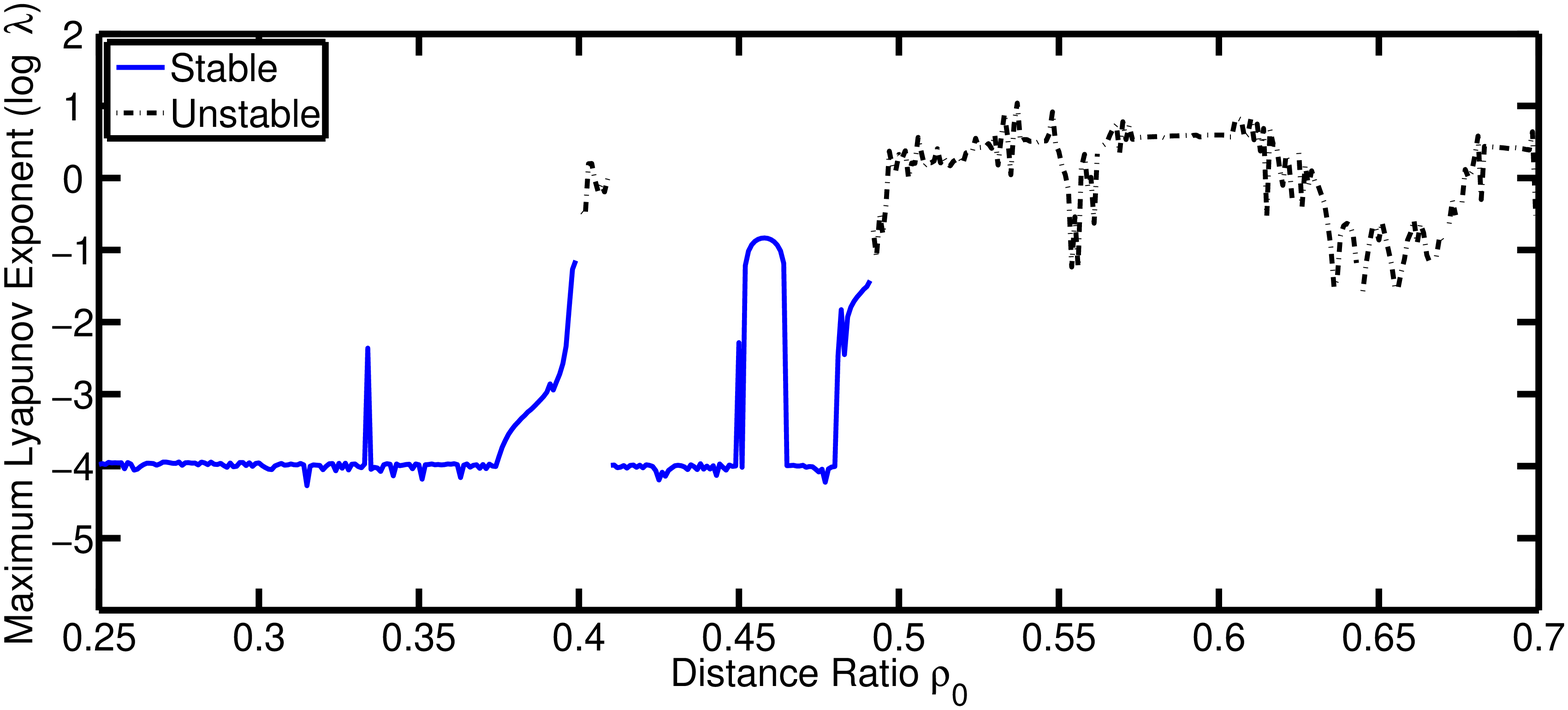,width=1.0\linewidth}

\caption{Same as Fig. 1 but for $\mu = 0.2$.}
\label{fig:2}       
\end{figure*}

\begin{figure*}[ht]
  \centering
  \epsfig{file=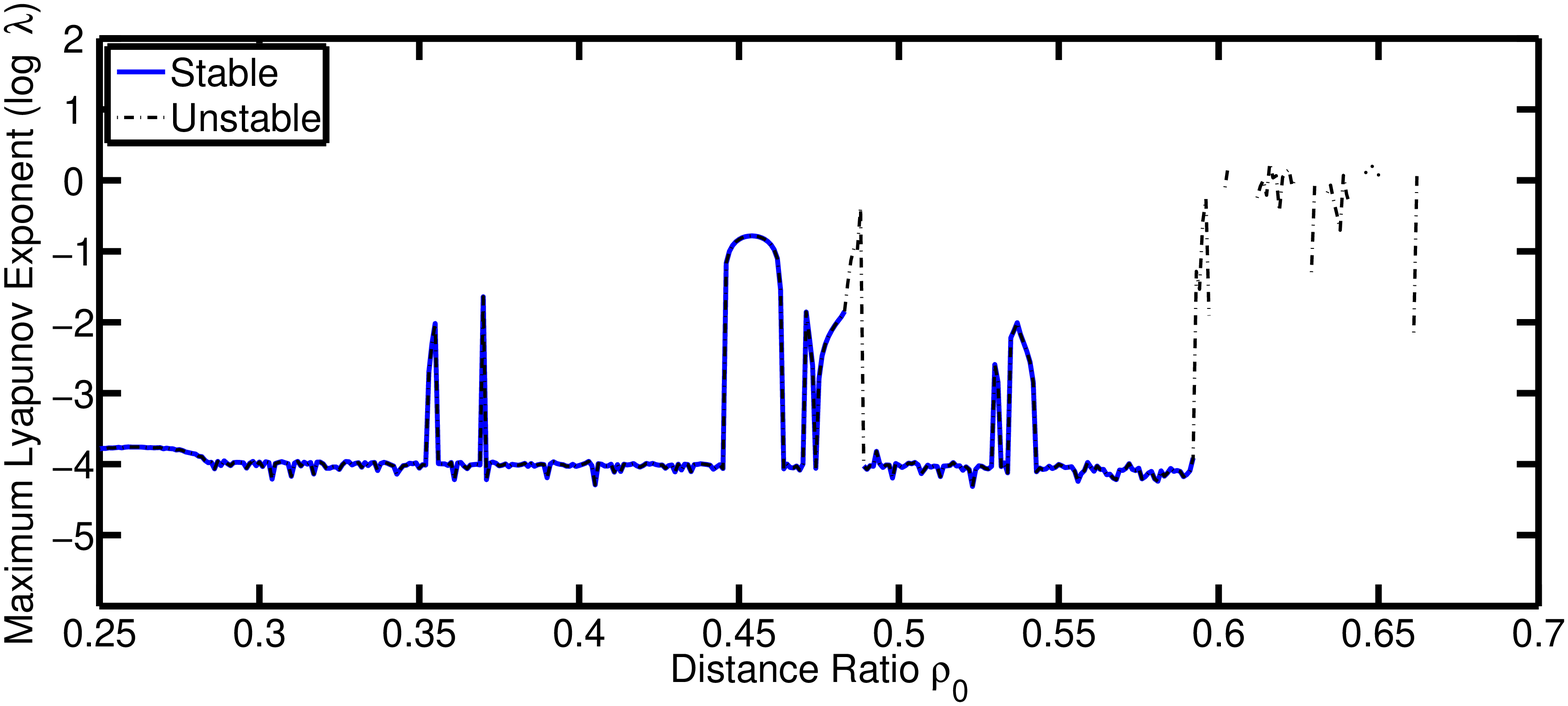,width=1.0\linewidth}
\caption{Same as Fig. 1 but for $\mu = 0.3$.}
\label{fig:3}       
\end{figure*}

In order to determine resonances, we utilize the method of Lyapunov exponents
\citep[e.g.,][]{nes98,nes02}.  In this method, we consider a constant mass ratio
$\mu$, calculate the spectrum of Lyapunov exponents, and monitor
 the maximum Lyapunov exponent with respect to the parameter $\rho_0$.  
Although we calculate the full spectrum of Lyapunov exponents, we 
limit this discussion to the maximum Lyapunov exponent for the reasons 
explained elsewhere \citep[e.g.,][]{qua11}.  The maximum Lyapunov 
exponent is used as an indicator of the maximum role that a particular resonance will 
have on the system.  This provides us with a way of inspecting how the
maximum Lyapunov exponent changes in response to small variations in 
the initial condition $\rho_0$.  Additionally, we use the fast Lyapunov 
indicator \citep*[e.g.,][]{fro97,leg01} to estimate the extent of weak 
chaotic or ordered motion.

After choosing a value of $\mu$ 
and performing calculations of the maximum Lyapunov exponent, we use
a FFT to produce a periodogram, and then take ratios of the periods in 
the periodogram to determine what kind of resonances (if any) are present.
Using the Lyapunov exponent with a periodogram provides a method of
quickly identifying regions of possible resonances rather than invoking
a brute force method of performing a FFT over all possible values of
$\rho_0$.  The ratios of the periods in the periodogram are taken with
respect to the period axis; however, the strength of the peaks
are not considered.  In the tables, we provide the periods of the
three strongest peaks in the periodogram and label them as Peak 1,
Peak 2, and Peak 3 in ascending order.  Resonance 1 denotes the
closest rational expression of the ratio of Peak 2 to Peak 1,
whereas Resonance 2 denotes the closest rational expression of Peak 3
to Peak 1.  More peaks may be possible in the periodograms; however,
the induced instability greatly overwhelms the necessity for investigating
them.  There are also elements in the tables without data (shown by ellipsis).
They indicate simulations where a Peak 3 could not be discriminated against the background;
therefore, a value for Resonance 2 could not be determined.
We study the range of $\rho_0$ beginning at $\rho_0 = 0.250$ and
ending at $\rho_0 = 0.700$ in increments of $0.001$ for a given value of $\mu$.


\begin{figure*}[ht]
  \centering
  \epsfig{file=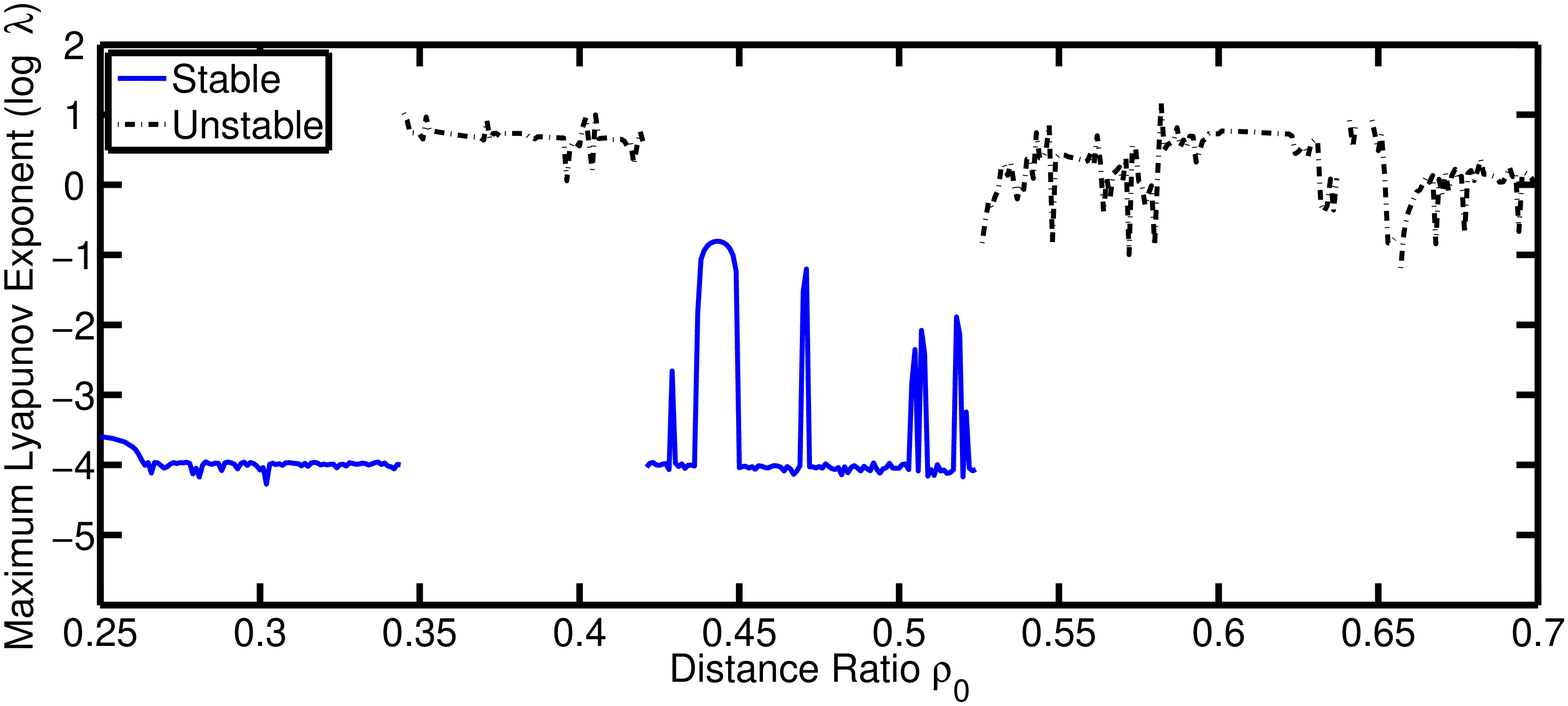,width=1.0\linewidth}
\caption{Same as Fig. 1 but for $\mu = 0.4$.}
\label{fig:4}       
\end{figure*}

\begin{figure*}[ht]
  \centering
  \epsfig{file=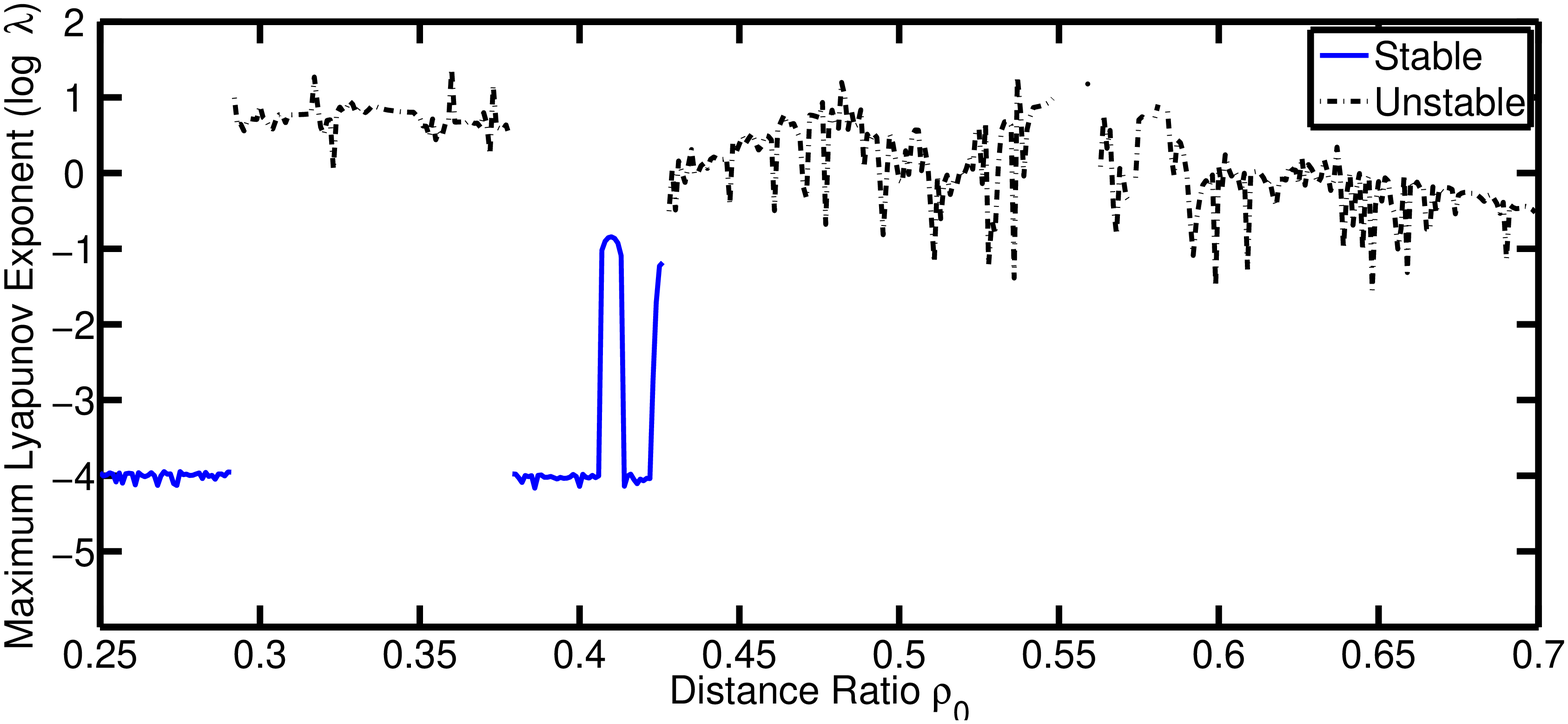,width=1.0\linewidth}
\caption{Same as Fig. 1 but for $\mu = 0.5$.}
\label{fig:5}       
\end{figure*}

\begin{table*}
\caption{Resonances for selected values of $\rho_0$ with an elevated 
maximum Lyapunov exponent given for $\mu = 0.1$.  The different orbital 
types are classified as Periodic, Quasi-Periodic, and Non-Periodic, and
are denoted by P, QP, and NP, respectively.}
\label{tab:table1}
\centering
\begin{tabular}{ c  c  c  c  c  c  c}
\hline
\noalign{\smallskip}
$\rho_0$ & Peak 1 & Peak 2 & Peak 3$^a$ & Resonance 1 & Resonance 2$^a$ & Orbital Type \\
\noalign{\smallskip}
\hline
\noalign{\smallskip}
0.394 & 0.3494 & 1.045 & \ldots & 3:1 & \ldots & P\\
0.457 & 0.4693 & 0.8827 & 1.001 & 15:8 & 17:8 & P\\
0.471 & 0.4832 & 0.9671 & \ldots & 2:1 & \ldots & P\\
\noalign{\smallskip}
\hline
\end{tabular}
\vspace{0.05in}
\begin{list}{}{}
\item[]$^a$Note: Elements without data indicate simulations that did not have a Peak 3 in the periodogram;
therefore, a value for Resonance 2 could not be determined.
\end{list}
\end{table*}

\begin{table*}

\caption{Same as Table 1 but for $\mu = 0.2$.}
\label{tab:table2}
\centering
\begin{tabular}{ c  c  c  c  c  c  c}
\hline
\noalign{\smallskip}
$\rho_0$ & Peak 1 & Peak 2 & Peak 3 & Resonance 1 & Resonance 2 & Orbital Type \\
\noalign{\smallskip}
\hline
\noalign{\smallskip}
0.334 & 0.2715 & 1.087 & \ldots & 4:1 & \ldots & P\\
0.391 & 0.3631 & 1.143 & \ldots & 22:7 & \ldots & QP\\
0.399 & 0.3756 & 1.127 & \ldots & 3:1 & \ldots & QP\\
0.450 & 0.4636 & 0.7716 & 1.162 & 5:3 & 5:2 & QP\\
0.458 & 0.4769 & 0.815 & 1.151 & 12:7 & 12:5 & QP\\
0.482 & 0.5153 & 1.031 & \ldots & 2:1 & \ldots & P\\
0.491 & 0.3457 & 0.5186 & 1.037 & 3:2 & 3:1 & QP\\
\noalign{\smallskip}
\hline
\end{tabular}

\end{table*}

\begin{table*}

\centering
\caption{Same as Table 1 but for $\mu = 0.3$.}
\label{tab:table3}
\begin{tabular}{ c  c  c  c  c  c  c}
\hline
\noalign{\smallskip}
$\rho_0$ & Peak 1 & Peak 2 & Peak 3 & Resonance 1 & Resonance 2 & Orbital Type \\
\noalign{\smallskip}
\hline
\noalign{\smallskip}
0.355 & 0.3267 & 1.297 & \ldots & 4:1 & \ldots & QP\\
0.370 & 0.3503 & 1.357 & \ldots & 31:8 & \ldots & QP\\
0.454 & 0.4743 & 1.448 & \ldots & 3:1 & \ldots & QP\\
0.471 & 0.4996 & 1.500 & \ldots & 3:1 & \ldots & P\\
0.483 & 0.5121 & 0.7680 & 1.537 & 3:2 & 3:1 & QP\\
0.488 & 0.5244 & 0.8034 & 1.511 & 3:2 & 23:8 & NP\\
0.493 & 0.5236 & 0.8278 & 1.495 & 14:3 & 14:5 & QP\\
0.530 & 0.5906 & 0.9875 & 1.472 & 5:3 & 5:2 & QP\\
0.537 & 0.6016 & 1.017 & 1.475 & 5:3 & 22:9 & QP\\
\noalign{\smallskip}
\hline
\end{tabular}

\end{table*}

\begin{table*}

\centering
\caption{Same as Table 1 but for $\mu = 0.4$.}
\label{tab:table4}
\begin{tabular}{ c  c  c  c  c  c  c}
\hline
\noalign{\smallskip}
$\rho_0$ & Peak 1 & Peak 2 & Peak 3 & Resonance 1 & Resonance 2 & Orbital Type \\
\noalign{\smallskip}
\hline
\noalign{\smallskip}
0.429 & 0.4456 & 1.995 & \ldots & 9:2 & \ldots & QP\\
0.444 & 0.4726 & 1.732 & \ldots & 11:3 & \ldots & QP\\
0.471 & 0.5236 & 1.580 & \ldots & 3:1 & \ldots & QP\\
0.505 & 0.2958 & 0.5918 & 1.480 & 2:1 & 5:1 & QP\\
0.507 & 0.2976 & 0.5953 & 1.473 & 2:1 & 5:1 & QP\\
0.518 & 0.2046 & 0.3070 & 0.614 & 3:2 & 3:1 & QP\\
0.521 & 0.2060 & 0.3090 & 0.618 & 3:2 & 3:1 & QP\\
\noalign{\smallskip}
\hline
\end{tabular}

\end{table*}

\begin{table*}

\centering
\caption{Same as Table 1 but for $\mu = 0.5$.}
\label{tab:table5}
\begin{tabular}{ c  c  c  c  c  c  c}
\hline
\noalign{\smallskip}
$\rho_0$ & Peak 1 & Peak 2 & Peak 3 & Resonance 1 & Resonance 2 & Orbital Type \\
\noalign{\smallskip}
\hline
\noalign{\smallskip}
0.410 & 0.232 & 0.4639 & \ldots & 2:1 & \ldots & P\\
0.426 & 0.251 & 0.5014 & 1.506 & 2:1 & 6:1 & QP\\
\noalign{\smallskip}
\hline
\end{tabular}

\end{table*}

\section{Results and Discussion} 

\subsection{Case studies}

We consider five different cases of fixed mass ratio $\mu$, starting at
$\mu = 0.1$ and proceeding in increments of $0.1$.  Each case presents a different aspect 
of the problem but together the combined results allow us to identify and 
classify resonances in the coplanar CR3BP for relatively large values of $\mu$.
For each case, we start with the parameter 
$\rho_0$ at the value $0.250$ that is increased in increments of $0.001$ up
to a final value of $\rho_0=0.700$.  We show one plot and one table for each
case of $\mu$.  In each plot, we distinguish between the stable cases, 
which continued for the full $10^5$ binary orbits of simulation time by a solid line, 
and the unstable cases by dashed lines; the latter ended early due to the ejection
of the planet from the system.  Our criteria 
for ending each simulation are based on checking the error in the Jacobi 
constant and the value of the maximum Lyapunov exponent where high values of either
imply a high probability
for $m_3$ to be ejected from the system \citep{qua11}.  We also pursued numerical comparisons
with an independent integration method to verify the integrity of the proposed outcome (see Section~\ref{sec:2}).
These separate criteria provide us with a method for checking
for instability and stability, as well as a measure for identifying chaos in
the system.  In the accompanying tables, we provide the values of $\rho_0$
where the maximum Lyapunov exponent is found to peak.

After the maximum Lyapunov exponent is calculated, we transform the system
pointwise from barycentric coordinates to Jacobi coordinates choosing the more massive body
as reference point.  We perform a FFT using the time series data for that specific $\rho_0$.  With 
this result, we obtain the Fourier spectrum and convert it to a 
corresponding periodogram.  The resonances are determined by taking 
the ratio of periods for the strongest peaks.  In the tables, we 
provide the computational results for the three strongest peaks,
if available, for that value of $\rho_0$.  The resonances are given as 
a ratio of the period of the small ($m_3$) mass to the period of 
the smaller binary ($m_2$) mass.  For example, 4:1 would 
tell us that the small mass orbits four times in a single orbit of 
the smaller binary mass.

The first case of $\mu = 0.1$ is depicted in Fig.~\ref{fig:1} with 
numerical results given in Table~\ref{tab:table1}.  This case 
illustrates a baseline of configurations where the second large ($m_2$) 
mass is not greatly perturbing the small ($m_3$) mass from $\rho_0 
= 0.250$ to $0.350$.  This provides a clear indication of where a
possible resonance is not substantial enough to alter the orbit.
However, the value of the maximum 
Lyapunov exponent peaks first at $\rho_0 = 0.394$.  This indicates 
that the second large mass is at a specific geometric position allowing it
to greatly perturb the motion of the small mass.  The same feature 
occurs again at $\rho_0 = 0.457$ and $0.471$.  Note that the resonances 
associated with each peak in Fig.~\ref{fig:1} are not all the same; see Table~\ref{tab:table1}.  
Moreover, we identify a large region of instability when $\rho_0$ is further increased, which
is consistent with previous investigations \citep{ebe10,qua11}.

The second case of $\mu = 0.2$ is depicted in Fig.~\ref{fig:2} and 
the relevant numerical values are presented in Table~\ref{tab:table2}.  
This case differs from the preceding one as now the first 
resonance peak occurs sooner at $\rho_0 = 0.334$.  This is expected 
because by increasing the value of $\mu$, the perturbing mass has
a greater ability to interact with the smaller mass to alter its orbit.
There exists a stability region from $\rho_0 = 0.250$ to $0.333$.  However,
the value of the maximum Lyapunov exponent peaks at seven different values
of $\rho_0$, which demonstrates the greater influence of the perturbing
mass $m_2$ compared to the case of $\mu = 0.1$.
Also, there are apparent gaps surrounding the instability regions.
Actually, they are due to our increment size in $\rho_0$ and,
therefore, these regions should be treated as continuous with appropriate
lines transitioning from one region to the next.  There is also
a resonance plateau in the range of $\rho_0 = 0.45$ to $0.48$.
This illustrates a region where the phenomena of resonance is widespread;
note there is a similar feature in the study of asteroids (i.e.,
the 3:1 resonance at 2.5 AU or $\rho_0 = 0.48$).

The third case of $\mu = 0.3$ is depicted in Fig.~\ref{fig:3} and 
the associated numerical results are given in Table~\ref{tab:table3}.  
Here the first resonance peak occurs at $\rho_0 = 0.355$.
It also exhibits a well-pronounced region where the resonance is not
substantial enough to alter the orbit of the small mass $m_3$ as shown in Fig.~\ref{fig:3}
from $\rho_0 = 0.250$ to $0.354$.  The value of the maximum Lyapunov
exponent peaks at $9$ different values of $\rho_0$ before reaching
the stability limit at $\rho_0 = 0.600$ as shown by \cite{ebe08}.
It is important to note that the number of resonance peaks has
increased thus far as the value of $\mu$ has increased.  Also this
case includes a peak in the maximum Lyapunov exponent, where the
system is unstable at $\rho_0 = 0.488$ leading to planetary ejection.
The transition to this instability is shown by the developing resonances
at $\rho_0 = 0.483$ in Table~\ref{tab:table3}.

The fourth case of $\mu = 0.4$ is depicted in Fig.~\ref{fig:4} with 
numerical results given in Table~\ref{tab:table4}.  This case differs
from the preceding trend by the first resonance peak; this peak is
delayed and occurs at $\rho_0 = 0.429$ after an instability island.  This is due to the
perturbing mass gaining an overwhelming amount of gravitational
influence over the small mass and creating an instability region
from $\rho_0 = 0.350$ to $0.400$.  Due to this change there are
fewer positions for resonance to occur, the resonances become denser with respect to $\rho_0$, and the value of the
maximum Lyapunov exponent peaks at $7$ different values of $\rho_0$.

The final case of $\mu = 0.5$ is depicted in Fig.~\ref{fig:5} and 
the relevant numerical results are given in Table~\ref{tab:table5}.  
This case differs as  
the first resonance peak is found to occur sooner at $\rho_0 = 0.410$.  The 
Copenhagen case shows the worst case scenario for the small mass 
because the tug of war between the large masses becomes the most 
extreme.  As a result there is only two narrow regions of stability.  
The value of the maximum Lyapunov exponent peaks at only two different 
values of $\rho_0$.  The first resonance characterizes a stable 
resonance, whereas the second is a resonance that shows the transition
from stability to instability via resonance overlap.

\subsection{Classification of resonances}

In Tables 1 to 5, we identified the existence of 2 possible resonances.  
We now discuss them separately starting with the results obtained for
Resonance 1.  These results show that the 2:1 resonance occurs
in all considered cases of $\mu$, except $\mu = 0.3$.  It is also shown
that the 2:1 resonance is the most common resonance in the coplanar
CR3BP in agreement with 
previous asteroidal investigations.  The 3:1 resonance 
is the second most common resonance as it occurs in all cases except 
$\mu = 0.5$.  A new and interesting result is that both the 3:2 and 5:3
resonances occur more often than the 4:1 resonance, which only 
occurs for $\mu = 0.2$ and $0.3$.

The overall picture obtained from the special cases of the 
values considered for $\mu$ is that the best chance for a resonance occurs at 
the intermediate value of $\mu = 0.3$.  In this case, we see that systems 
start with a resonance reflecting a large ratio of the small mass period 
to the large mass period.  Then as the parameter $\rho_0$ increases, so 
does the value of the ratio.  After crossing the smoothest peak 
at $\rho_0 = 0.454$, the secondary resonances are formed; these resonances 
are labeled as Resonance 2 in Tables 1 to 5 and once they appear 
their existence may imply instability caused by resonance overlap.  
Among these resonances, the 3:1 resonance is the most dominant but
new resonances such as 5:2 also occur, which is present for $\mu = 0.2$ 
and $0.3$, and 5:1 occurring for $\mu = 0.4$.  An interesting result is that
only the 3:1 resonance occurs also as a secondary resonance and that none
of the other resonances labeled as Resonance 1 also occur as Resonance 2.

\begin{figure*}[ht]
  \centering
  \subfloat[]{\epsfig{file=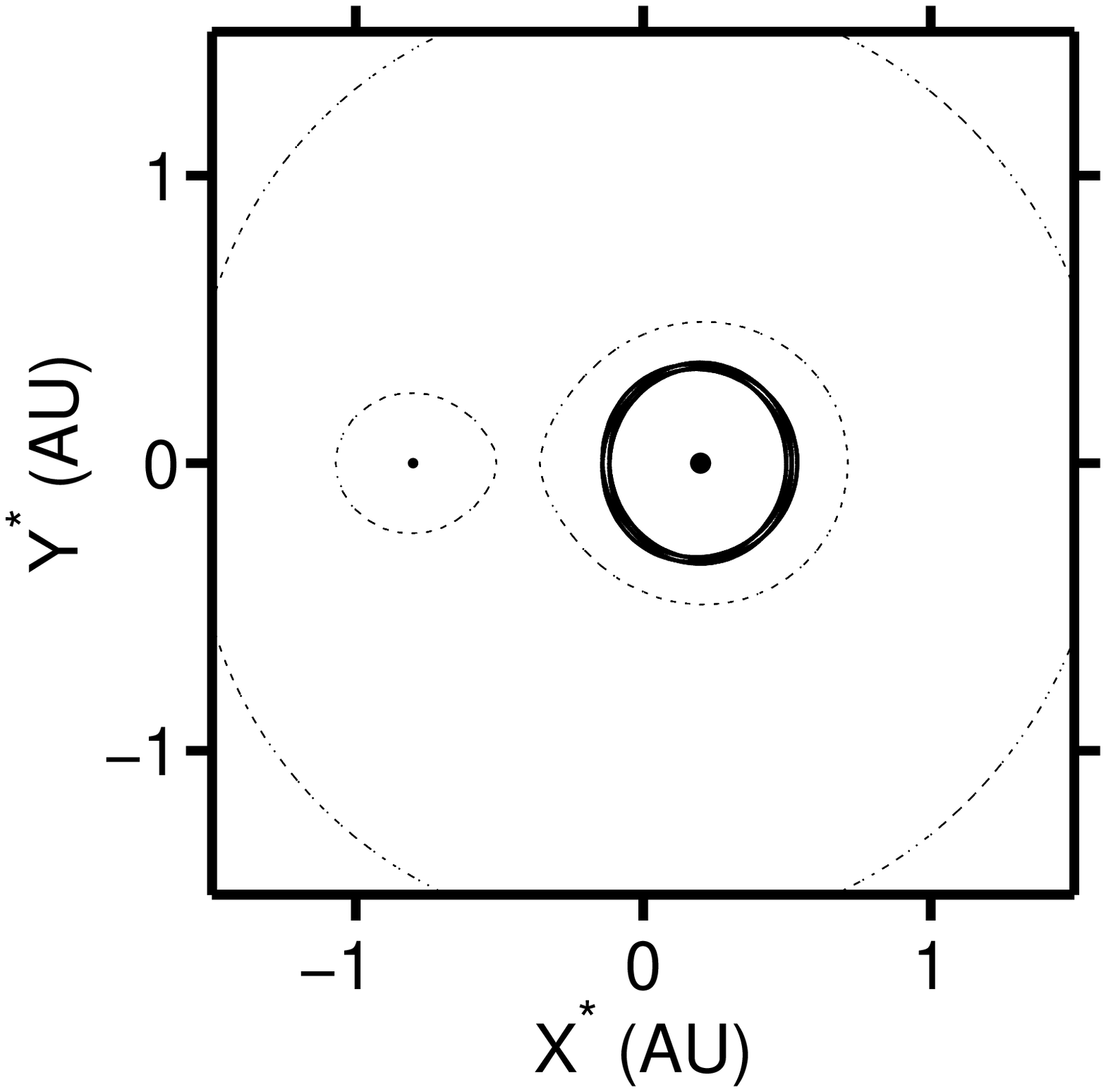,width=0.33\linewidth}} 
  \subfloat[]{\epsfig{file=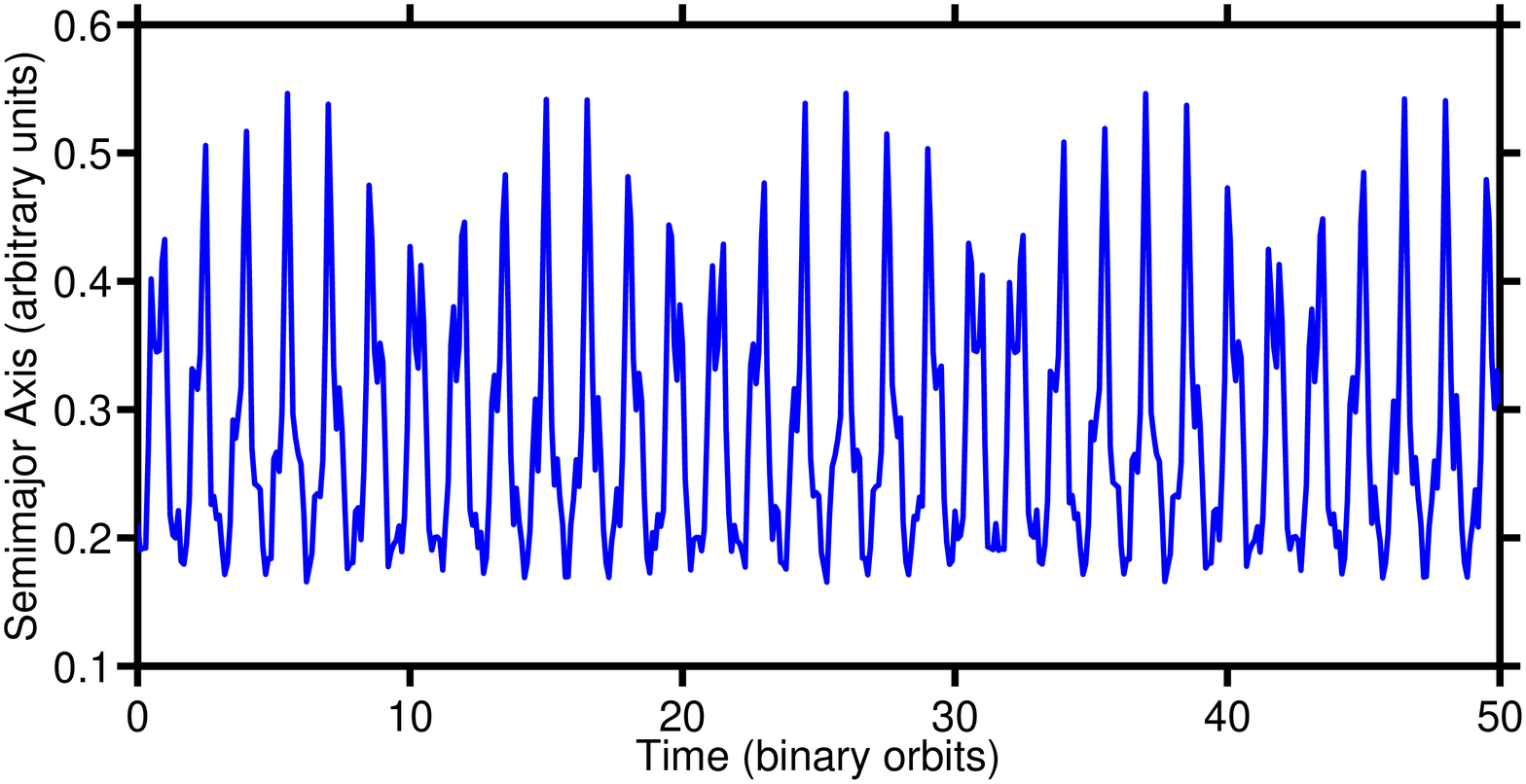,width= 0.5\linewidth,height=0.3\linewidth}} 
  \subfloat[]{\epsfig{file=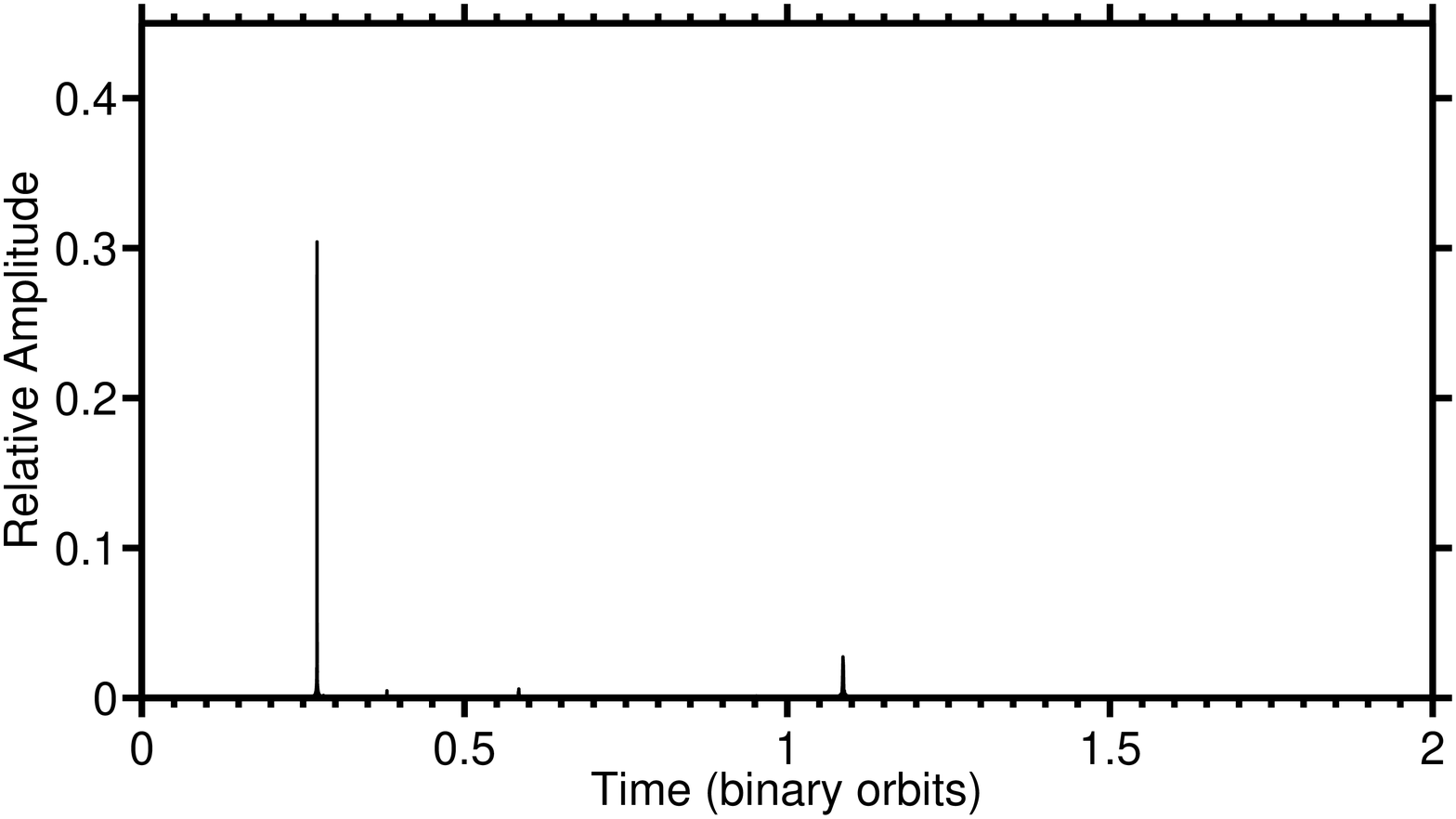,width= 0.3\linewidth,height=0.3\linewidth}}
\caption{Case study showing the results for $\mu = 0.2$ and $\rho = 0.334$.
Panel (a) shows the orbit of a planet in a rotating coordinates system,
(b) shows the osculating semimajor axis for the first 50 binary orbits, and
(c) shows the Fourier periodogram to determine the possible resonances.}
\label{fig:6}       
\end{figure*}

\begin{figure*}[ht]
  \centering
  \subfloat[]{\epsfig{file=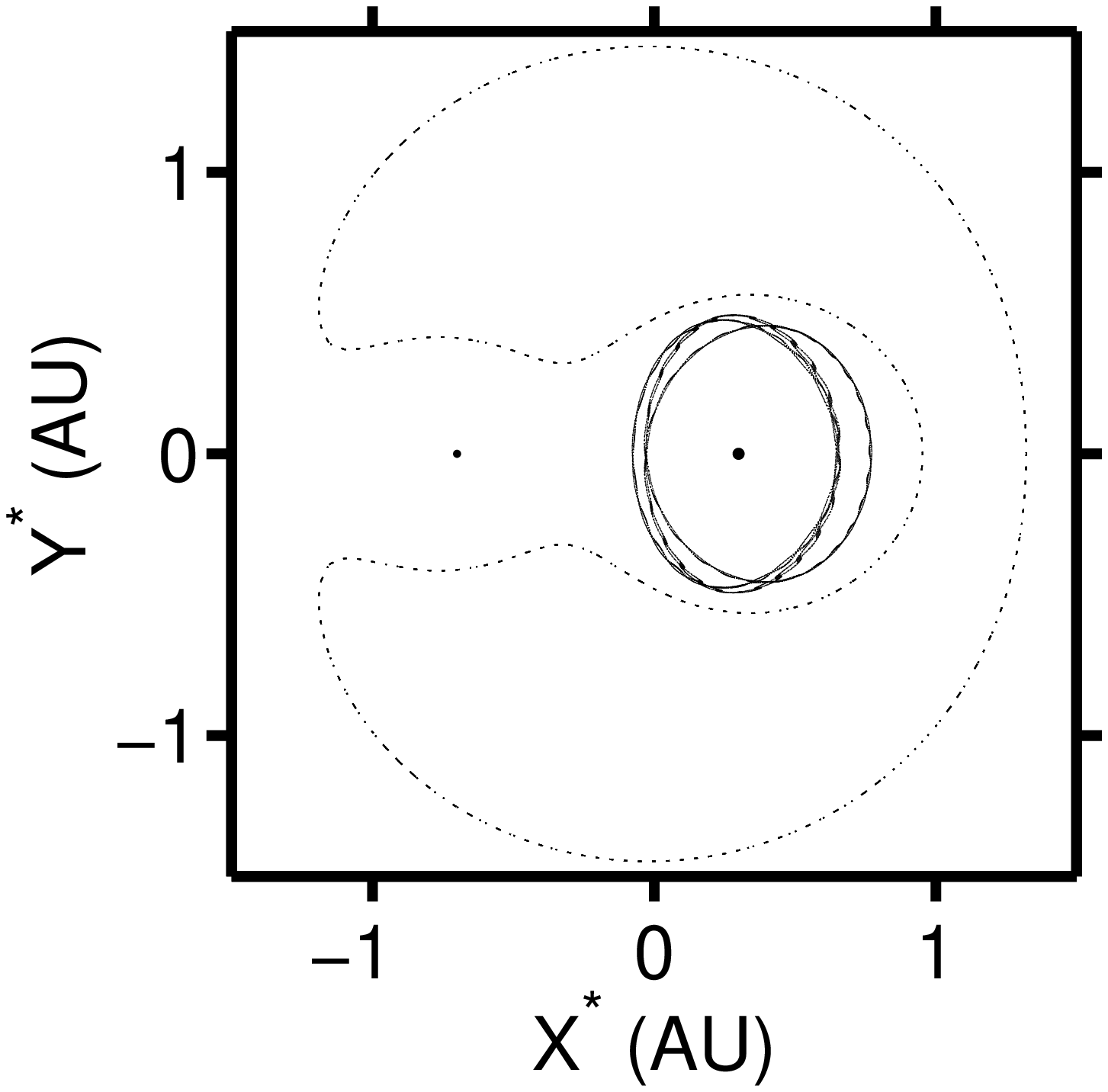,width=0.3\linewidth}} 
  \subfloat[]{\epsfig{file=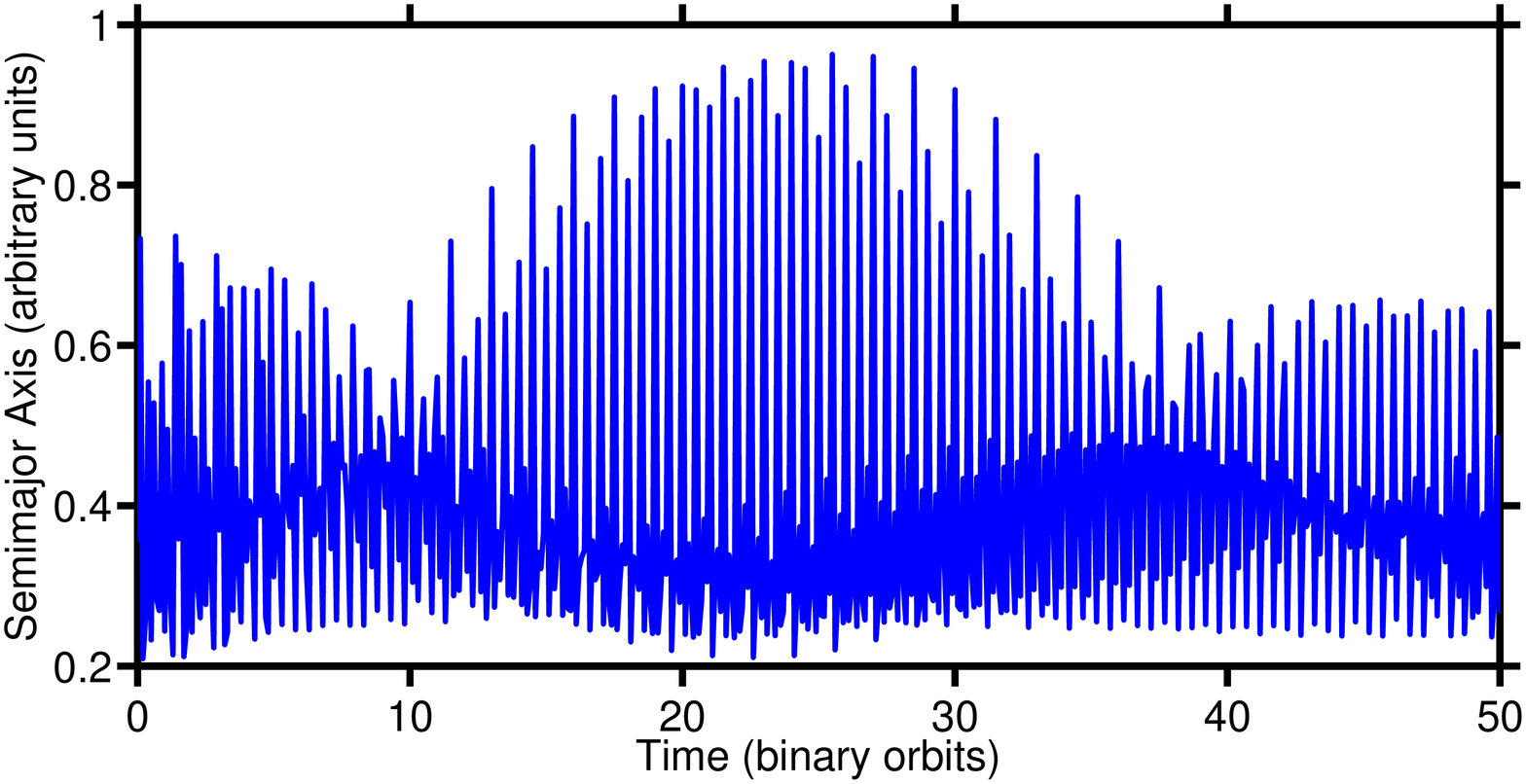,width= 0.5\linewidth,height=0.3\linewidth}} 
  \subfloat[]{\epsfig{file=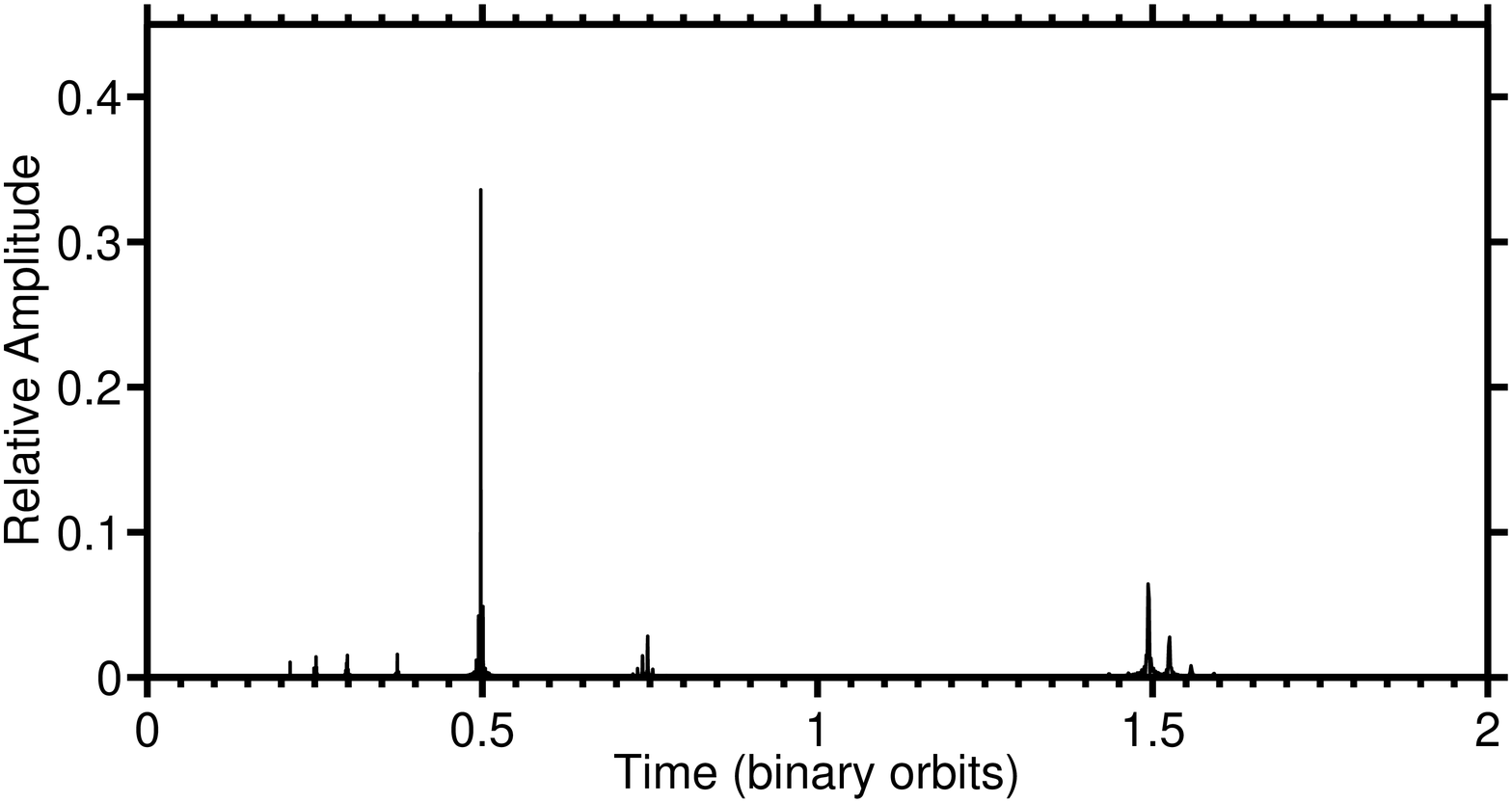,width= 0.3\linewidth,height=0.3\linewidth}}
\caption{Case study showing the results for $\mu = 0.3$ and $\rho = 0.471$.
Panel (a) shows the orbit of a planet in a rotating coordinates system,
(b) shows the osculating semimajor axis for the first 50 binary orbits, and
(c) shows the Fourier periodogram to determine the possible resonances.}
\label{fig:7}       
\end{figure*}

\begin{figure*}[ht]
  \centering
  \subfloat[]{\epsfig{file=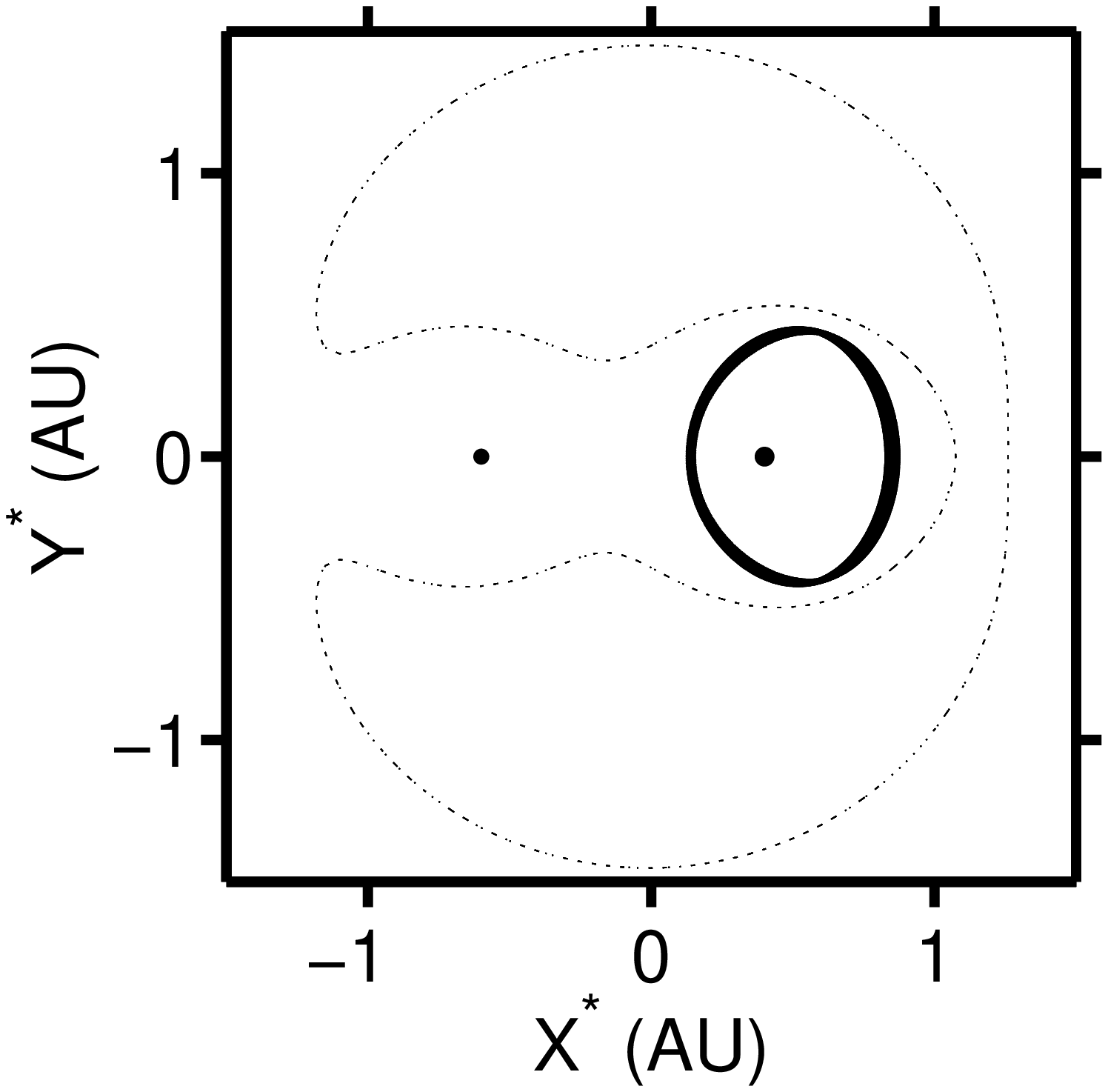,width=0.3\linewidth}} 
  \subfloat[]{\epsfig{file=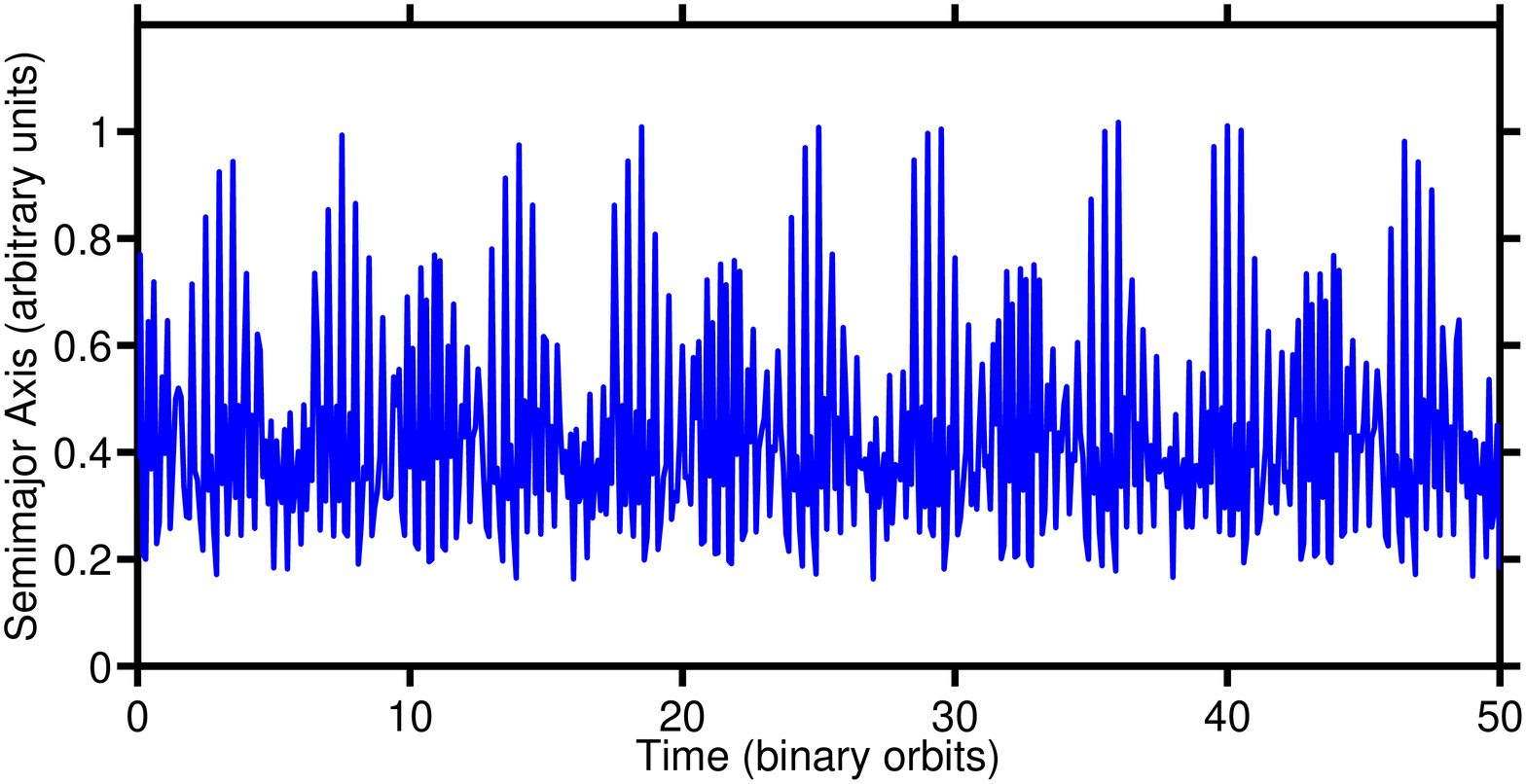,width= 0.5\linewidth,height=0.31\linewidth}} 
  \subfloat[]{\epsfig{file=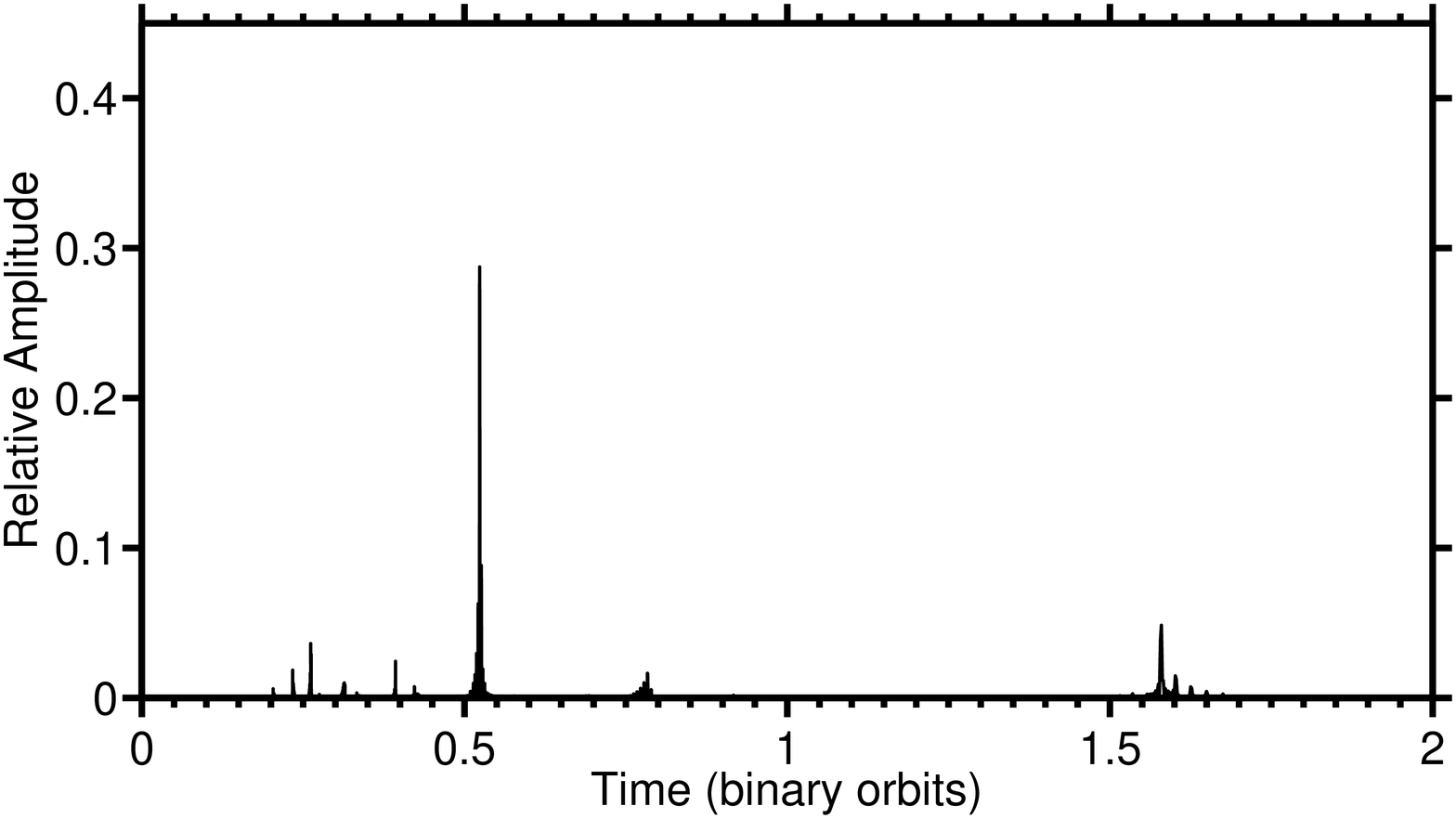,width= 0.3\linewidth,height=0.31\linewidth}}
\caption{Case study showing the results for $\mu = 0.4$ and $\rho = 0.471$.
Panel (a) shows the orbit of a planet in a rotating coordinates system,
(b) shows the osculating semimajor axis for the first 50 binary orbits, and
(c) shows the Fourier periodogram to determine the possible resonances.}
\label{fig:8}       
\end{figure*}

\begin{figure*}[ht]
  \centering
  \subfloat[]{\epsfig{file=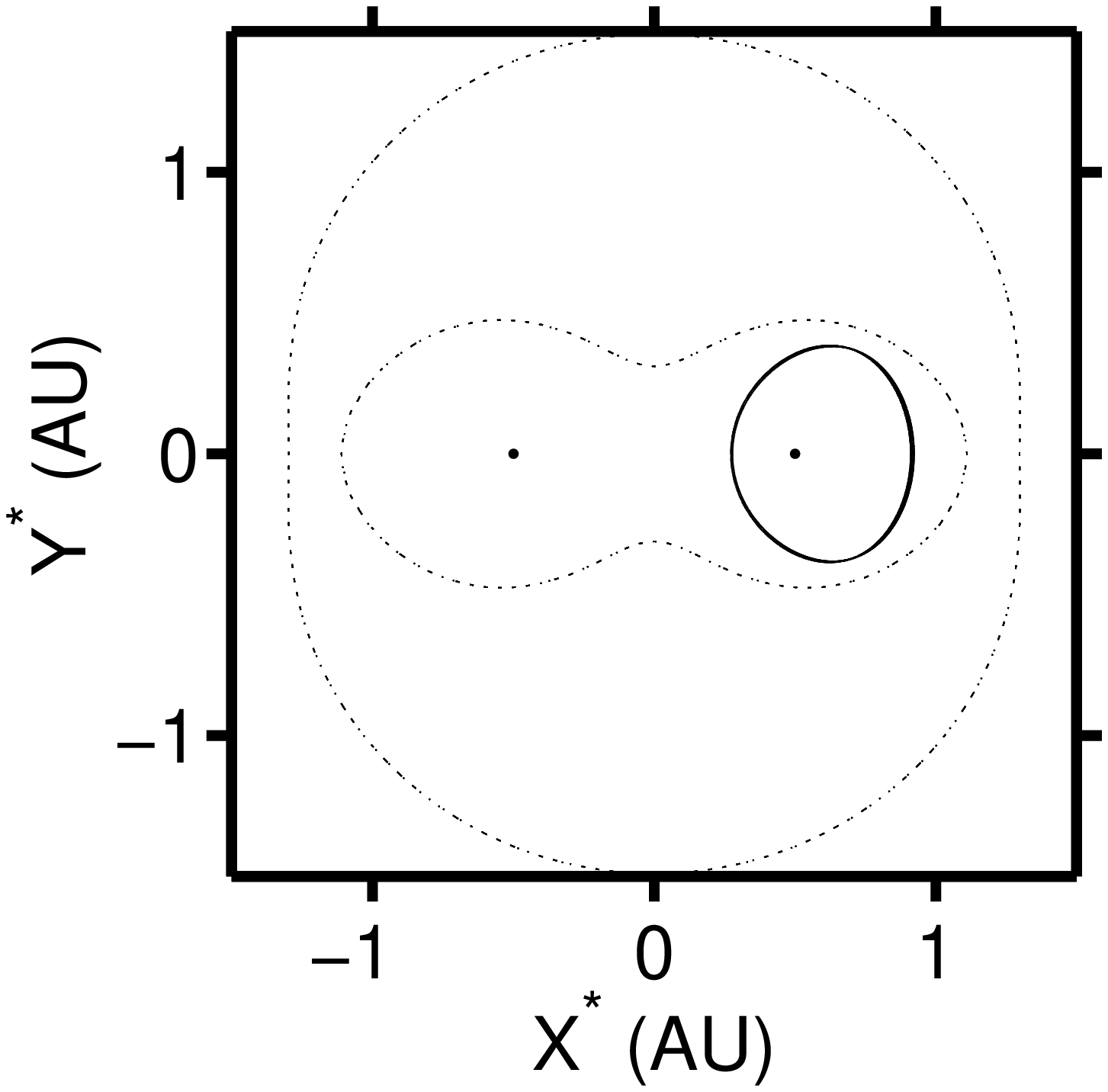,width=0.3\linewidth}} 
  \subfloat[]{\epsfig{file=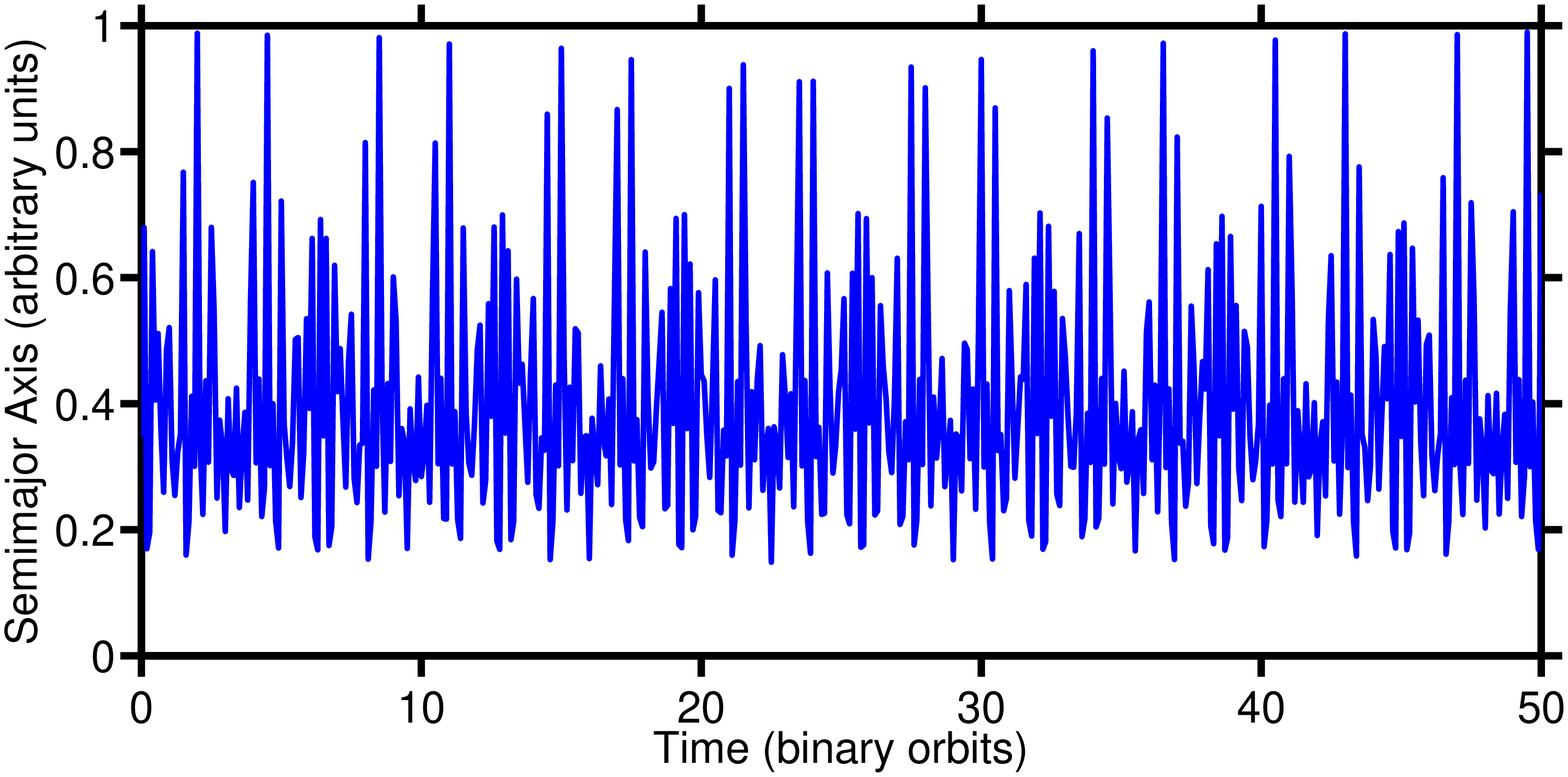,width= 0.5\linewidth,height=0.31\linewidth}} 
  \subfloat[]{\epsfig{file=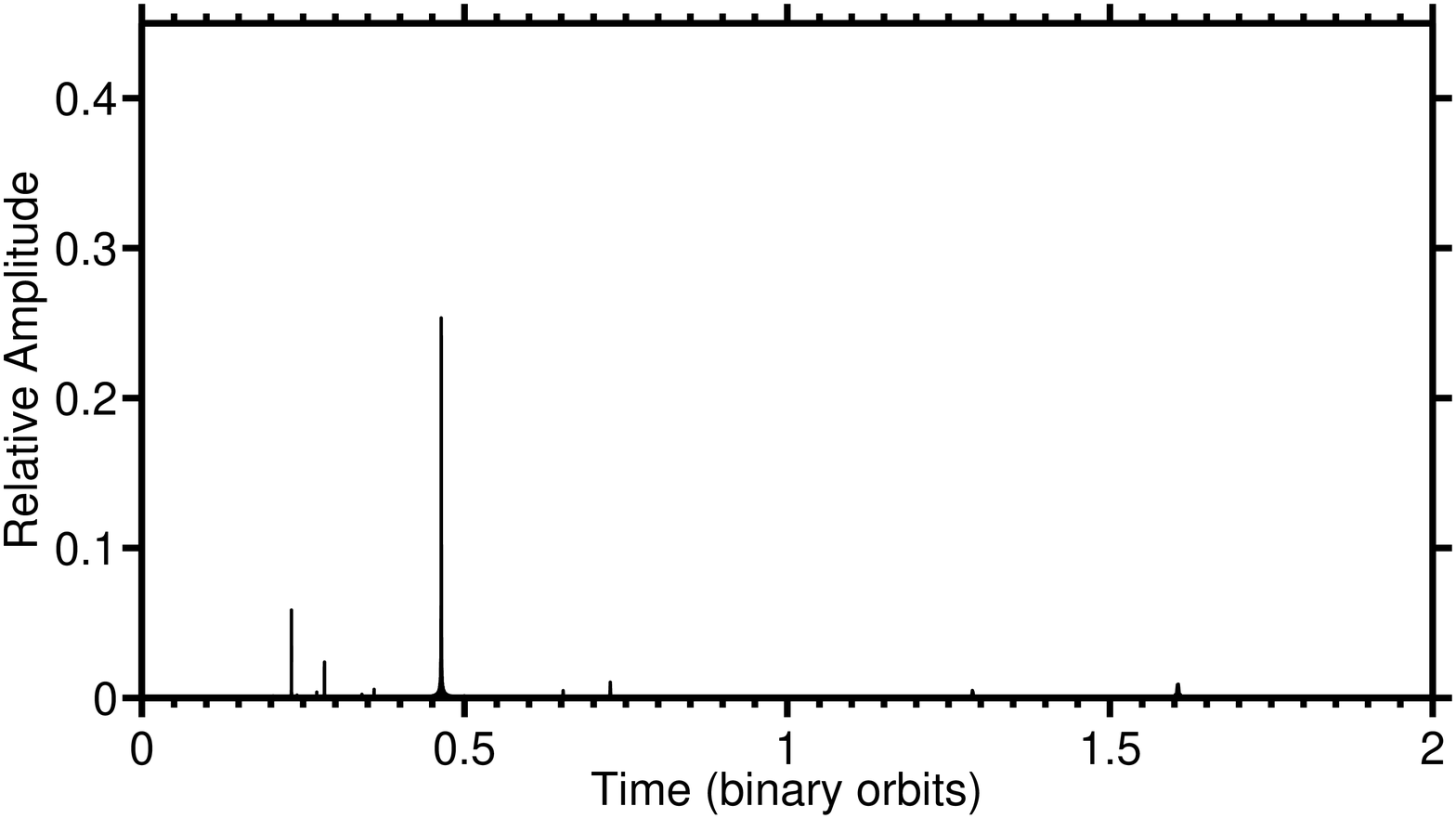,width= 0.3\linewidth,height=0.31\linewidth}}
\caption{Case study showing the results for $\mu = 0.5$ and $\rho = 0.410$.
Panel (a) shows the orbit of a planet in a rotating coordinates system,
(b) shows the osculating semimajor axis for the first 50 binary orbits, and
(c) shows the Fourier periodogram to determine the possible resonances.}
\label{fig:9}       
\end{figure*}

Additionally, we can classify resonances based on the behaviour of the 
orbits.  In Tables~\ref{tab:table1} to \ref{tab:table5}, we show 
classifications based on orbit diagrams in rotating coordinates
(see Figs. \ref{fig:6} to \ref{fig:9}a) and the osculating semimajor
axis of the small mass (see Figs. \ref{fig:6} to \ref{fig:9}b).  The 
three classes are periodic, quasi-periodic, and non-periodic,  
denoted by P, QP, and NP, respectively.  Periodic 
orbits are those that are well-defined (i.e., minimal deviation from the 
path traced by previous orbits).  They are closed, show periodic variations in the 
osculating semimajor axis, and show two dominant peaks in the corresponding periodogram
(see Figs. \ref{fig:6}c, \ref{fig:7}c, and \ref{fig:9}c).  Quasi-periodic orbits are 
those that precess or fill an annulus with an inner and outer boundary.
Moreover, quasi-periodic orbits exhibit a more complicated osculating semimajor axis
as well as a periodogram with several peaks (see Figs. \ref{fig:7}b,c).
Non-periodic orbits are those that do not show a regular pattern and are thus 
most likely to be unstable.  Also non-periodic cases show the greatest
number of peaks in their periodograms.  Of the cases presented, a
low value of $\mu$ yields a greater probability for periodic orbits.
As the mass ratio $\mu$ increases, the probability for quasi-periodic 
orbits increases due to increasing influence of the perturbative mass and 
duration of the perturbation.  However, there are some periodic orbits 
that exist for large values of $\mu$.  They occur near the smooth 
plateau that forms in each of the figures.



\section{Summary and conclusions} \label{sec:4}

The case studies of the coplanar CR3BP presented in this paper were used to identify
and classify resonances in the coplanar CR3BP.  Our method of finding resonances is 
based on the concept of Lyapunov exponents, or more specifically, it uses the
maximum Lyapunov exponent.  Based on the latter, we were able to determine 
where the resonances occurred for a fixed value of mass ratio $\mu$.  The conditions 
for uncovering these resonances were determined by the value of the maximum 
Lyapunov exponent at the end of each simulation.  The fact that the value of 
the maximum Lyapunov exponent peaks at specific values of distance ratio
$\rho_0$ provides a strong indicator for the positions of the resonances.
Using this method, we have shown that stability is ensured for high values of
resonance (i.e., high integer ratios) where only a single resonance is present.  

We identified the primary and secondary resonances in the coplanar CR3BP as Resonance 1
and 2 (see Tables 1 to 5), have shown that the 2:1 resonance occurs 
most frequently as Resonance 1 in the considered systems, and have shown that 
the 3:1 resonance is the next most common.  An interesting result is that
there are two other resonances, which are 3:2 and 5:3, and that they occur 
more often than the 4:1 resonance.  Among the resonances labeled as 
Resonance 2, the 3:1 resonance is the most dominant but there are also 
new resonances such as 5:2 and 5:1.  Moreover, we found that the 3:1
resonance is the only one that also occurs as a secondary resonance, 
and that none of the other resonances identified by us as Resonance 1
also occur as Resonance 2.  The existence of the resonances labeled as 
Resonance 2 is important as it may imply the onset of instability 
caused by the resonance overlap.

We also classified the resonances in the coplanar CR3BP based on the behaviour of the orbits.
The following three different classes were used: periodic (P), quasi-periodic (QP),
and non-periodic (NP); see \cite{ebe10} and \cite{qua11} for previous usages of
these classes in conjunction with proposed criteria for the onset of orbital
instability.  Among the considered cases, a low value for $\mu$ and $\rho_0$ entailed a
greater probability for periodic orbits to occur.  Then, the probability for
quasi-periodic orbits increases as $\mu$ increases.  However, we also discovered
some periodic orbits that exist for large values of $\mu$.
 
The resonances reported in this paper are consistent with the 
previously established resonances for the Solar System, specifically, in the
asteroid belt with its dominant 2:1 resonance \citep{fer94}.  
They used digital filtering and Lyapunov characteristic exponents to 
determine stochasticity of the eccentricity.  This is consistent with our usage 
of Lyapunov exponents as we took an alternate approach, which is varying the 
mass ratio instead of the eccentricity.  Our approach provides general results 
for the cases where $\mu$ is considerably greater than the value of $\mu$ in
Sun--Jupiter-type systems but where the eccentricity is small.
Although our results have been obtained for the special case of the coplanar CR3BP, 
we expect that it will be possible to augment our findings to planets in general
stellar binary systems.  Desired generalizations should include studies of the
ER3BP \citep[e.g.,][]{pil02,sze08} as well as applications to
planets in previously discovered star--planet systems.

\acknowledgements
This work has been supported by the U.S. Department of Education under GAANN
Grant No. P200A090284 (B.~Q.), the Alexander von Humboldt Foundation
(Z.~E.~M.) and the SETI institute (M.~C.).

\newpage

\end{document}